%% file: main.tex
\documentclass[conference]{IEEEtran}
\IEEEoverridecommandlockouts

\usepackage{cite}
\usepackage{amsmath,amssymb,amsfonts}
\usepackage{algorithmic}
\usepackage{graphicx}
\usepackage{textcomp}
\usepackage{xcolor}
\usepackage{color}
\usepackage[acronym]{glossaries}
\usepackage{float} 
\usepackage{url}
\usepackage{hyperref}
\usepackage{tikz}

\makeglossaries
\input{acronyms.tex}

\def\BibTeX{{\rm B\kern-.05em{\sc i\kern-.025em b}\kern-.08em
    T\kern-.1667em\lower.7ex\hbox{E}\kern-.125emX}}
\newcommand\submittedtext{
\footnotesize This work has been accepted for presentation at IEEE Quantum Week 2025: IEEE International Conference on Quantum Computing and Engineering (QCE).}
\newcommand\submittednotice{
\begin{tikzpicture}[remember picture,overlay]
\node[anchor=south,yshift=6pt] at (current page.south) {\fbox{\parbox{\dimexpr1\textwidth-\fboxsep-\fboxrule\relax}{\submittedtext}}};
\end{tikzpicture}
}
\renewcommand\fbox{\fcolorbox{black}{white}}
\begin{document}

\title{Leveraging Diffusion Models for Parameterized Quantum Circuit Generation\\
\submittednotice
\thanks{This work was partially funded by the BMWK project EniQmA (01MQ22007A).}
}

\author{\IEEEauthorblockN{1\textsuperscript{st} Daniel Barta}
\IEEEauthorblockA{\textit{Technische Universitat Berlin} \\
\textit{Fraunhofer FOKUS}\\
Berlin, Germany \\
daniel.barta@fokus.fraunhofer.de}
\and
\IEEEauthorblockN{2\textsuperscript{nd} Darya Martyniuk}
\IEEEauthorblockA{\textit{Fraunhofer FOKUS} \\
\textit{}\\
Berlin, Germany \\
darya.martyniuk@fokus.fraunhofer.de}
\and
\IEEEauthorblockN{3\textsuperscript{rd} Johannes Jung}
\IEEEauthorblockA{\textit{Fraunhofer FOKUS} \\
\textit{}\\
Berlin, Germany \\
johannes.jung@fokus.fraunhofer.de}
\and
\IEEEauthorblockN{4\textsuperscript{th} Adrian Paschke}
\IEEEauthorblockA{\textit{Freie Universitat Berlin} \\
\textit{Fraunhofer FOKUS}\\
Berlin, Germany \\
adrian.paschke@fokus.fraunhofer.de}
}

\maketitle

\begin{abstract}
Quantum computing holds immense potential, yet its practical success depends on multiple factors, including advances in quantum circuit design.
In this paper, we introduce a generative approach based on denoising diffusion models~(DMs) to synthesize parameterized quantum circuits~(PQCs). Extending the recent diffusion model pipeline of~\cite{Frrutter2023QuantumCS}, our model effectively conditions the synthesis process, enabling the simultaneous generation of circuit architectures and their continuous gate parameters. We demonstrate our approach in synthesizing PQCs optimized for generating high-fidelity Greenberger–Horne–Zeilinger (GHZ) states and achieving high accuracy in quantum machine learning (QML) classification tasks. 
Our results indicate a strong generalization across varying gate sets and scaling qubit counts, highlighting the versatility and computational efficiency of diffusion-based methods. This work illustrates the potential of generative models as a powerful tool for accelerating and optimizing the design of~PQCs, supporting the development of more practical and scalable quantum applications.
\end{abstract}

\begin{IEEEkeywords}
Quantum Circuit Synthesis, Parameterized Quantum Circuits, Generative Models, Quantum Computing, Quantum Architecture Search, Diffusion Models.
\end{IEEEkeywords}

\section{Introduction}
The promise of quantum computing hinges on the efficient design of quantum circuits executable on~\gls{nisq} hardware~\cite{Preskill2018QuantumCI}. This is challenging due to hardware constraints like limited qubit counts and restricted gate sets~\cite{anagolum2024elivagar, patel2024curriculum, wang2022quantumnas}. 
The design of an appropriate ansatz for a given task~–~be it approximating a target unitary, solving an optimization problem or learning from data for a~\gls{qml} task~–~is challenging and typically relies on human expertise or brute-force search. 
In recent years, there has been growing interest in leveraging heuristic optimization and \gls{ml} techniques to automate quantum circuit design. 
Prior approaches include evolutionary strategies~\cite{chivilikhin2020mog, sunkel2023ga4qco, creevey2023gasp, ARUFE2022101030}, reinforcement learning~\cite{Ostaszewski2019QuantumCS, Moro_2021, patel2024curriculum, ostaszewski2021reinforcementlearningoptimizationvariational}, Bayesian optimization~\cite{duong2022quantum}, and adaptive algorithms~\cite{grimsley2019adaptive, zhu2022adaptive}.
While these methods can discover novel circuit designs, they typically require repeated evaluations of quantum metrics or cost functions during training~\cite{Arrazola2018MachineLM},~\cite{ Zhang2021NeuralPB}. 
This evaluation process becomes intractable for large systems and trainable \gls{pqcs}, creating a significant computational bottleneck for automated circuit synthesis methods~\cite{sarra2023discoveringquantumcircuitcomponents}.
Moreover, techniques trained on specific quantum tasks or circuit layouts often show poor generalization to new scenarios, as they tend to encode task-specific features rather than learning transferable principles~\cite{Preti_2024}.
To address these challenges,~\cite{Frrutter2023QuantumCS} introduces the idea of applying denoising diffusion models~(DM) to quantum circuit synthesis. 
The proposed text-conditioned diffusion pipeline generates gate-based circuits for a given target objective without explicitly simulating quantum dynamics during training. 
While this work established DMs as a promising tool for quantum circuit synthesis, an important open challenge is handling continuous gate parameters. 
Commonly used rotation gates, such as single-qubit or controlled rotations, are parameterized by continuous values and significantly impact the functionality of~\gls{pqcs}.

In this paper, we introduce an extended DM-based pipeline that integrates continuous-valued gate parameters alongside discrete gate choices into the generation process. 
We train the proposed model architecture on extensive datasets of quantum circuits with varying gate sets, designed for state preparation and QML classification tasks. 
Experimental results demonstrate that the described methodology effectively bypasses the computationally expensive parameter training phase during the search, addressing a key limitation of many existing approaches to the automated design of~\gls{pqcs}, and shows a strong ability to generalize to new scenarios.
At inference time, the model, guided by a textual prompt, produces~\gls{pqcs} that are immediately deployable on the target quantum device, subject to specified constraints.

The remainder of this paper is structured as follows.
Section~\ref{sec:background} discusses related work. Section~\ref{sec:methods} presents our approach for synthesizing \gls{pqcs}, detailing the diffusion-based pipeline and training procedure. Experimental results are presented in Section~\ref{sec:results}, with an analysis of potentials and limitations in Section~\ref{sec:discussion}. Finally, Section~\ref{sec:conclusion} concludes the paper and provides an outlook on future research.

\section{Related Work}\label{sec:background}

The approach introduced in this work extends~\cite{Frrutter2023QuantumCS}, which treats quantum circuits as structured data samples and trains a generative denoising DM to generate new circuit designs. DMs generate synthetic data by iteratively denoising random noise samples through a learned reverse diffusion procedure.
Introduced by~\cite{sohldickstein2015deepunsupervisedlearningusing}, the fundamental principle involves adding Gaussian noise to training samples as a forward Markovian diffusion process, then training a neural network to reverse the noising procedure and approximate the data distribution’s score function.
After training, data generation begins from a pure noise sample, and the trained model iteratively predicts and removes noise, step by step, resulting in samples resembling those from the training dataset.
DMs employ denoising autoencoder architectures implemented through neural networks, typically structured to predict either the original noise-free data or the noise itself. 
Recent advances, such as latent diffusion~\cite{rombach2022high} and~\gls{cfg}, have made DMs a leading approach in conditional generative modeling tasks across numerous domains, including image and video generation~\cite{rombach2022high, ho2022video}, audio synthesis~\cite{kong2020diffwave}, and protein structure prediction~\cite{jing2023eigenfold}. 
Extending these advances,~\cite{Frrutter2023QuantumCS} explored the potential of DMs beyond classical domains and demonstrated their effectiveness in the context of quantum circuit synthesis.
Specifically, the authors applied  a denoising DM to two quantum computing tasks: (1) entanglement generation, where the model produced circuits that realize specific multi-qubit entangled states, and (2) unitary compilation, where the model generated circuits approximating random target unitaries using constrained gate sets. The~DM excelled in generating valid, novel circuits and enabled advanced capabilities such as masking and editing parts of a circuit post hoc without retraining.
Notably, the approach eliminates the need for classical simulations to compute cost functions comparing generated circuits to target unitaries. 
Ref.~\cite{chen2025uditqc} proposes an architectural enhancement by replacing the standard U-Net used for noise prediction with a U-Net-style Diffusion Transformer.
A recent work~\cite{shen2023prepareansatzvqediffusion} applied a DM to generate ansätze for \gls{vqe}, with a focus on chemistry problems. 
The model was trained on a corpus of unitary coupled cluster~(UCC) ansatz circuits encoded into images and successfully proposed new circuits that achieved comparable ground-state energies.
However, the techniques proposed in earlier studies represent gates using fixed-dimensional vectors drawn from a finite embedding set~\cite{fösel2021quantumcircuitoptimizationdeep}, limiting its ability to express parametrized quantum operations. 
In this work, we extend the methodology to enable the generation of~\gls{pqcs} with continuous gate parameters.

Beyond diffusion-based models, this work also relates more broadly to the field of \textit{quantum architecture search~(QAS)} research, where the goal is to automatically generate~\gls{pqcs} tailored to specific tasks and hardware constraints~\cite{martyniuk2024quantum}. 
While the most approaches proposed for QAS treat the design of PQCs as a two-step process~\cite{Preti_2024, patel2024curriculum, Zhang2021NeuralPB, duong2022quantum},~i.e.,~first selecting discrete gate sequences and then using classical optimizers to determine continuous parameters, several recent methods aim to learn both circuit structures and parameterizations simultaneously, optimizing over both discrete gates and continuous parameters in a unified pipeline.
Ref.~\cite{altmann2024challenges} incorporates
the parameter optimization task into reinforcement learning settings, enabling continuous actions. The generative quantum eigensolver~(GQE) proposed in~\cite{nakaji2024generative} is a transformer-based approach for generating \gls{pqcs}. This approach is designed for the chemical problem of searching for the
ground state of the electronic structure Hamiltonians of several molecules. Similarly to our approach, it integrates continuous
optimizable gate parameters into a classical generative model.
However, unlike the transformer-based model in GQE, we employ a DM for \gls{pqcs} generation, which iteratively refines random noise samples to generate circuits, offering advantages in generating diverse quantum circuits with novel structures.

\section{Methods}\label{sec:methods}
In the following, we present the proposed DM-based pipeline for the generation of~\gls{pqcs}, detailing the circuit encoding strategy, model architecture, and training procedure.


    


\begin{figure*}[htbp]
    \centering
    \includegraphics[
    width=\textwidth,
    trim=1.2cm 1.1cm 1cm 0.7cm,   
    clip 
    ]{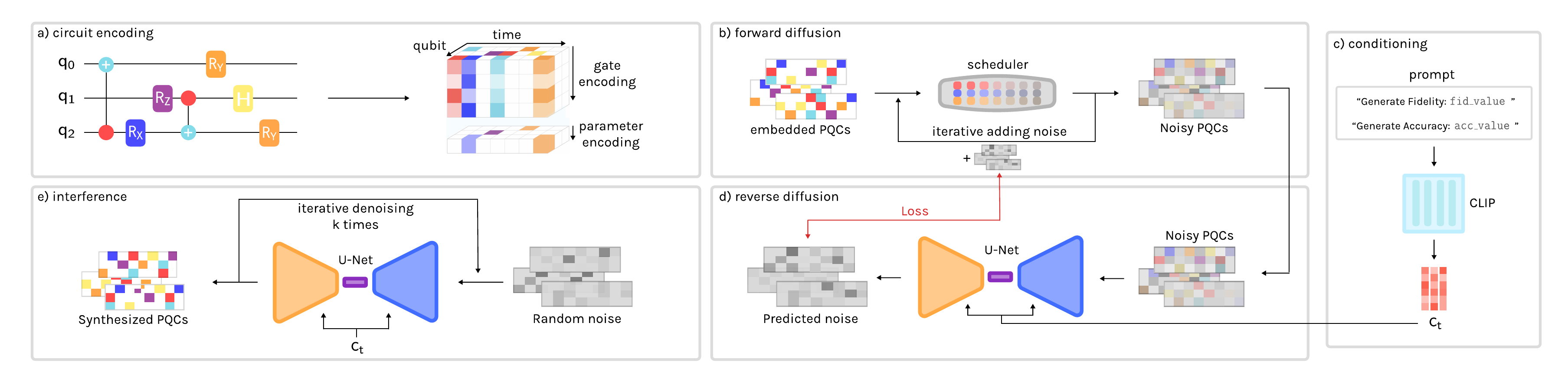}
       \caption{\textbf{Overview of our diffusion-based PQC synthesis pipeline, inspired by \cite{Frrutter2023QuantumCS}} 
        (a) \textbf{Circuit Encoding:} A circuit is converted into a tensor by concatenating the embeddings of discrete gates and continuous parameters (e.g., rotation angles). 
        (b) \textbf{Forward Diffusion:} Gaussian noise is progressively injected into the tensor according to a predefined scheduler, emulating a corruption process.
        (c) \textbf{Textual Conditioning:} A pretrained CLIP encoder converts task-specific prompts into embeddings. 
        (d) \textbf{Reverse Diffusion:} A U-Net with cross-attention receives the noisy tensor and text embedding, iteratively removing noise to restore gate identities and parameters. 
        (e) \textbf{Inference and Decoding:} At inference, random noise is denoised under the given prompt, and the final tensor is mapped back into a readable quantum circuit.
        }

    \label{fig:complete_pipe}
\end{figure*}

\subsection{Circuit Encoding}\label{sec:encoding}

To train our denoising DM, we represent each PQC using two structured tensors, as depicted in Fig.~\ref{fig:complete_pipe}~(a). 
The first tensor is three-dimensional, where one dimension corresponds to qubit indices~$N$, another to discrete time steps~$T$, and the third dimension encodes discrete gate identities~$d_c$. 
For each discrete time step~$t_i \in T$, only one gate per qubit is allowed. 
If no gate acts on a qubit at a particular timestep (i.e., an idle operation), we assign a special no-op token.
Multi–qubit gates are represented by role–specific tokens placed on all participating qubits in the same time slot (e.g.,\ a 'cx' acting on qubits $i$ and $j$, where '-' (minus) represents target and '+' (plus) represents control).
To convert this discrete representation into a continuous embedding suitable for the DM, each discrete gate type is transformed into a fixed-length embedding vector determined before training. This embedding process results in a continuous tensor representation of gates across qubits and time steps.
To handle gates with continuous parameters, e.g., a rotation gate 'rx($\theta$)', we introduce a second tensor with dimensions corresponding to qubit indices, discrete time steps, and the maximal number of normalized parameters used by any gate in the circuit~$d_p$. Note that currently, our approach supports parameter embeddings with~$d_p\leq1$. In this parameter tensor, we encode the numerical values of gate parameters, placing a no-op token for gates without parameters. By concatenating these two tensors, each gate operation for every qubit and timestep is uniquely encoded by its discrete type embedding and its continuous parameter. As a result, identical gates with identical parameters map to identical points in this combined embedding space, whereas gates differing in type or parameter values map distinctly. This method naturally extends to gates with multiple parameters.
During decoding, the combined tensor is split into a circuit matrix and a parameter matrix. For each qubit–timestep position in the circuit matrix, we use cosine similarity to match the generated gate embedding to its closest counterpart among the known gate embeddings. The corresponding parameter matrix is then applied only when a gate requires parameters, using the remaining $d_p$ dimensions of the vector as a parameter encoding. In principle, if a gate supports multiple parameters (not present in our current gate sets), one could introduce an inverse mapping or train an auxiliary decoder network to handle this more complex parameter space. If the chosen gate is non-parameterized (e.g., an 'h' gate), the parameter subvector is simply ignored.

\subsection{Model Architecture}\label{sec:architecture}
In this work, we extend the quantum circuit synthesis pipeline described by~\cite{Frrutter2023QuantumCS}. Specifically, we introduce textual conditioning via a pre-trained ~\gls{clip} text encoder~-~originally trained to align visual and textual representations~-~to guide the DM towards generating~\gls{pqcs} optimized for specific tasks, i.e., quantum state fidelity or classification accuracy maximization (Fig.~\ref{fig:complete_pipe}~(c)). Task-specific textual prompts, for instance \texttt{"Generate GHZ fidelity: }$\mathsf{fid\_value}$" or \texttt{"Generate Accuracy: }$\mathsf{acc\_value}$", are encoded into continuous embeddings to condition the diffusion process.

The core of our diffusion model utilizes a U-Net architecture, widely adopted in state-of-the-art diffusion-based image synthesis. 
The model accepts as input a noisy quantum circuit tensor of dimensions $(d_c + d_p)\times N\times T$, representing embedding dimensions for gates and parameters, qubits and time steps, respectively.
It outputs a noise-residual tensor of matching shape. To enable textual conditioning, we augment the standard U-Net structure with cross-attention layers, integrating~\gls{clip}-generated textual embeddings. This design explicitly conditions the generated circuits on task-specific textual prompts, eliminating the need for explicit labels of fidelity or accuracy, which the model implicitly learns from training data alone. 

\subsection{Training Strategy}\label{sec:training}

Our DM is trained by simulating a forward diffusion process on quantum circuit data (Fig.~\ref{fig:complete_pipe}~(b)) and learning its reverse denoising counterpart in a manner consistent with \gls{ddpm}~\cite{ho2020denoisingdiffusionprobabilisticmodels} procedures (Fig.~\ref{fig:complete_pipe}~(d)). Following the approach in~\cite{Frrutter2023QuantumCS}, we parameterize the model as an $\epsilon$-predictor: at each diffusion timestep $t$, the model learns to directly predict the noise $\epsilon_t$ in a noisy tensor $x_t = \sqrt{\bar{\alpha}_t}x_0 + \sqrt{1-\bar{\alpha}_t}\epsilon_t$, where $\epsilon_t \sim \mathcal{N}(0, I)$ (i.e., $\epsilon_t$ is drawn from a Gaussian distribution with zero mean and identity covariance), $x_0$ is a sample from the training dataset with arbitrary distribution $q(x_0)$, and $\bar{\alpha}_t$ is the cumulative product of the variance schedule (i.e., $\bar{\alpha}_t = \prod_{i=0}^t (1-\beta_i)$). The U-Net predicts this noise as $\hat{\epsilon}_\theta(x_t, t, \texttt{prompt})$, conditioned on the task prompt. Model parameters are optimized to minimize the mean-squared error loss:
\[
\mathcal{L} = \mathbb{E}_{x_0, \epsilon, t} \left[\left\|\epsilon - \hat{\epsilon}_\theta(x_t, t, \texttt{prompt})\right\|^2 \right],
\]
ensuring the network learns to denoise samples at all diffusion steps~\cite{Frrutter2023QuantumCS}.

The U-Net architecture comprises an encoder-decoder convolutional network with residual blocks and transformer-based spatial attention layers. It processes a single unified tensor of size $(d_c + d_p)\times N\times T$, where each slice along the embedding dimension encodes a gate identity (or no-op) and, if applicable, its continuous parameter. We incorporate the diffusion timestep~$t$ via a sinusoidal positional encoding added within the residual blocks, allowing the network to learn how to invert the noise process at every step.

Conditioning on textual prompts (e.g., "\texttt{Generate GHZ fidelity: 1.0000}") is done through cross-attention layers, where the prompt is embedded by a ~\gls{clip} encoder (output size \(77 \times 512\)). For numerical stability, fidelity prompts are rounded to four decimal places, while QML accuracy prompts use one decimal place. We adopt this approach because small changes in prompt phrasing can induce noticeable variations in the generated circuits; thus, prompt sensitivity is mitigated by discretizing the input conditions. During training, with a 10\% probability, the prompt is replaced by an empty token. This “unconditional” path enables~\gls{cfg} at inference, whereby we can smoothly interpolate between conditional and unconditional predictions~\cite{ho2022classifierfreediffusionguidance}. We train for \(T=1000\) diffusion steps using Adam optimizer with a one-cycle learning-rate schedule.

At inference time, we draw Gaussian noise as an initial circuit encoding and iteratively denoise it using the trained U-Net (Fig.~\ref{fig:complete_pipe}~(e)). When~\gls{cfg} is enabled, we selectively mix conditional and unconditional predictions to control how strictly the model follows the prompt. The final denoised tensor is decoded into a human-readable quantum circuit: we pick the closest discrete gate embedding at each position and apply the predicted continuous parameter (if required) to obtain the final PQC.

\section{Results}\label{sec:results}
In this section, we present the experimental results, evaluating the proposed approach and analyzing its performance across two common quantum computing tasks: quantum state preparation and a QML task focused on linear classification.

\subsection{Dataset}
A key element of our approach is the construction of large training datasets consisting of quantum circuits, which the diffusion model leverages to learn the circuit performance distribution. 
To address our two evaluation tasks, namely quantum state preparation and classification accuracy optimization, we utilize the existing SQuASH\footnote{Available at~\url{https://github.com/SQuASH-bench/SQuASH}} dataset~\cite{SQuASH}. Specifically, for the \textit{quantum state preparation}, the dataset includes quantum circuits designed to generate a $3$-qubit GHZ state within a predefined approximation error. 
The data were collected from various search strategies used to generate circuits for the task, including random search, reinforcement learning, and an evolutionary approach. 
To demonstrate the model’s adaptability to different hardware constraints, we define two distinct data subsets, each characterized by a different gate alphabet.
The first subset incorporates PQCs constructed using the gate set $gs1$~=~\{'cx', 'h', 'rx', 'ry', 'rz', 'id'\}. The second subset uses~$gs2$~=~\{'cz', 'id', 'rx', 'rz', 'rzz', 'sx', 'x'\}, which corresponds to the native gate set of the IBM quantum processing unit~(QPU)~\textit{'ibm\_fez'}.

We observed that strategically balancing the GHZ-task datasets with respect to both fidelity distribution and circuit length yielded substantial improvements in model performance. Specifically, in $gs1$, 75\% of the circuits had high fidelity ($>0.9$) without additional length balancing, while the remaining 25\% had low fidelity ($\le 0.9$) evenly distributed across all circuit lengths. In $gs2$, 90\% of the circuits had high fidelity without length balancing, and the other 10\% had low fidelity evenly distributed across the full range of circuit lengths. Furthermore, we restricted the number of gates per circuit to 3–16 for $gs1$ and 3–24 for $gs2$. Empirically, this targeted balancing approach notably improved training stability and enhanced the quality of generated quantum circuits.


For the \textit{machine learning setting}, SQuASH provides a dedicated set of quantum circuits along with their performance on the \textit{LinearlySeparable}
\footnote{ \url{https://github.com/XanaduAI/qml-benchmarks/blob/main/src/qml_benchmarks/data/linearly_separable.py}}
dataset introduced in~\cite{bowles2024better}, which is a linear classification task. 
The dataset consists of 300 samples, each with $d = 8$ features, and a predefined margin of $m = 1$ between the two classes. To ensure reproducibility, we fix the random seed to 274 when generating the data.
As in the previous task, we define a specific gate set, \{’cx’, ’h’, ’rx’, ’ry’, ’swap’, ’crx’, ’cry’\}, and strategically balance the dataset in terms of classification accuracy and circuit length, which significantly improves training performance.  
Similar to the quantum state preparation, we constructed the dataset so that 50\% of circuits had accuracy below 0.7, evenly balanced across various circuit lengths, while the remaining 50\% consisted of high-accuracy circuits ($\geq0.7$) without further length balancing. All circuits were constrained to 3-24 gates. Empirical results confirm that this targeted dataset composition notably enhances model performance.



\subsection{GHZ Task}
\begin{figure}[!t]
\centering
\begin{minipage}[c]{0.49\linewidth}
\centering
\includegraphics[width=\linewidth]{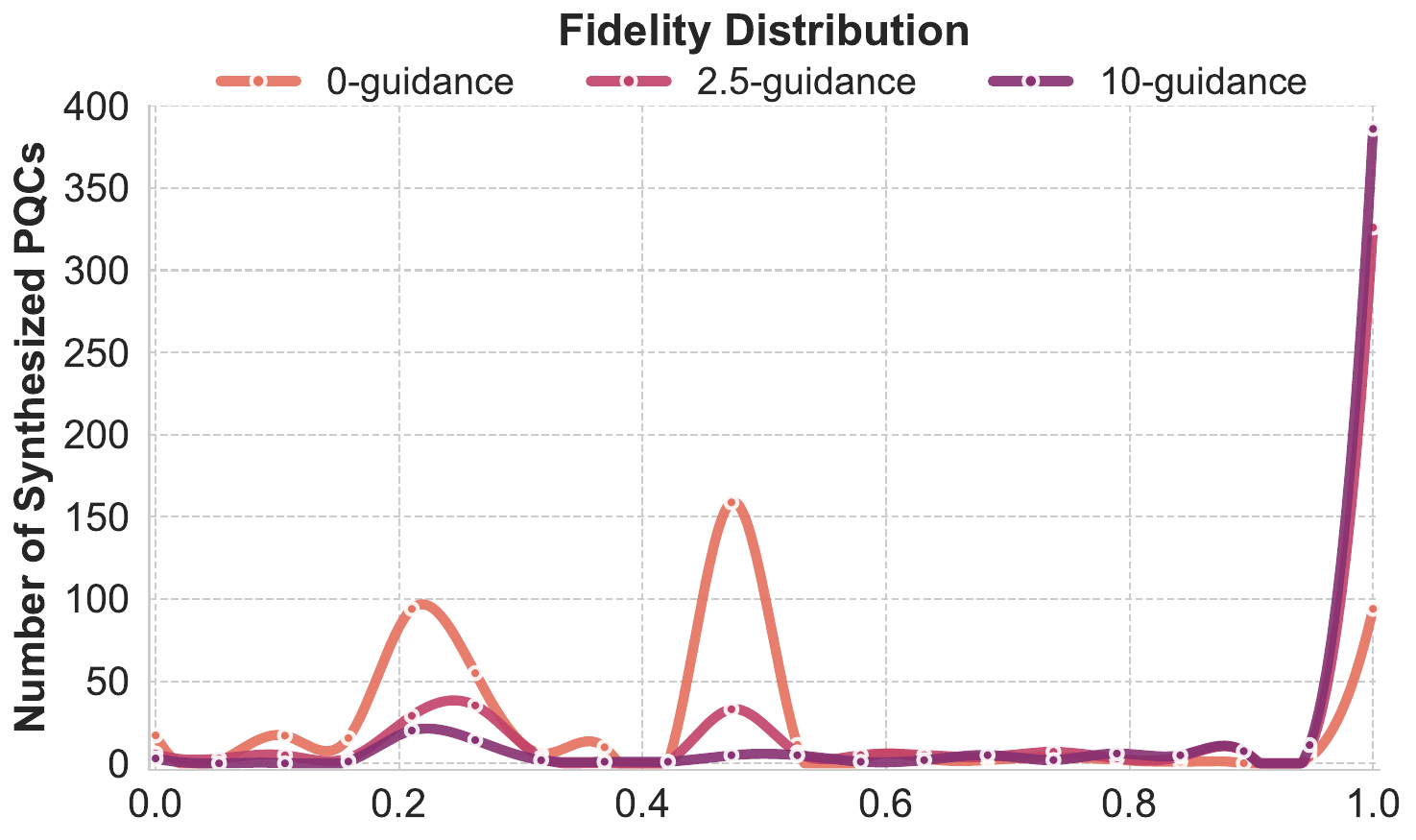}
\end{minipage}
\begin{minipage}[c]{0.49\linewidth}
\centering
\includegraphics[width=\linewidth]{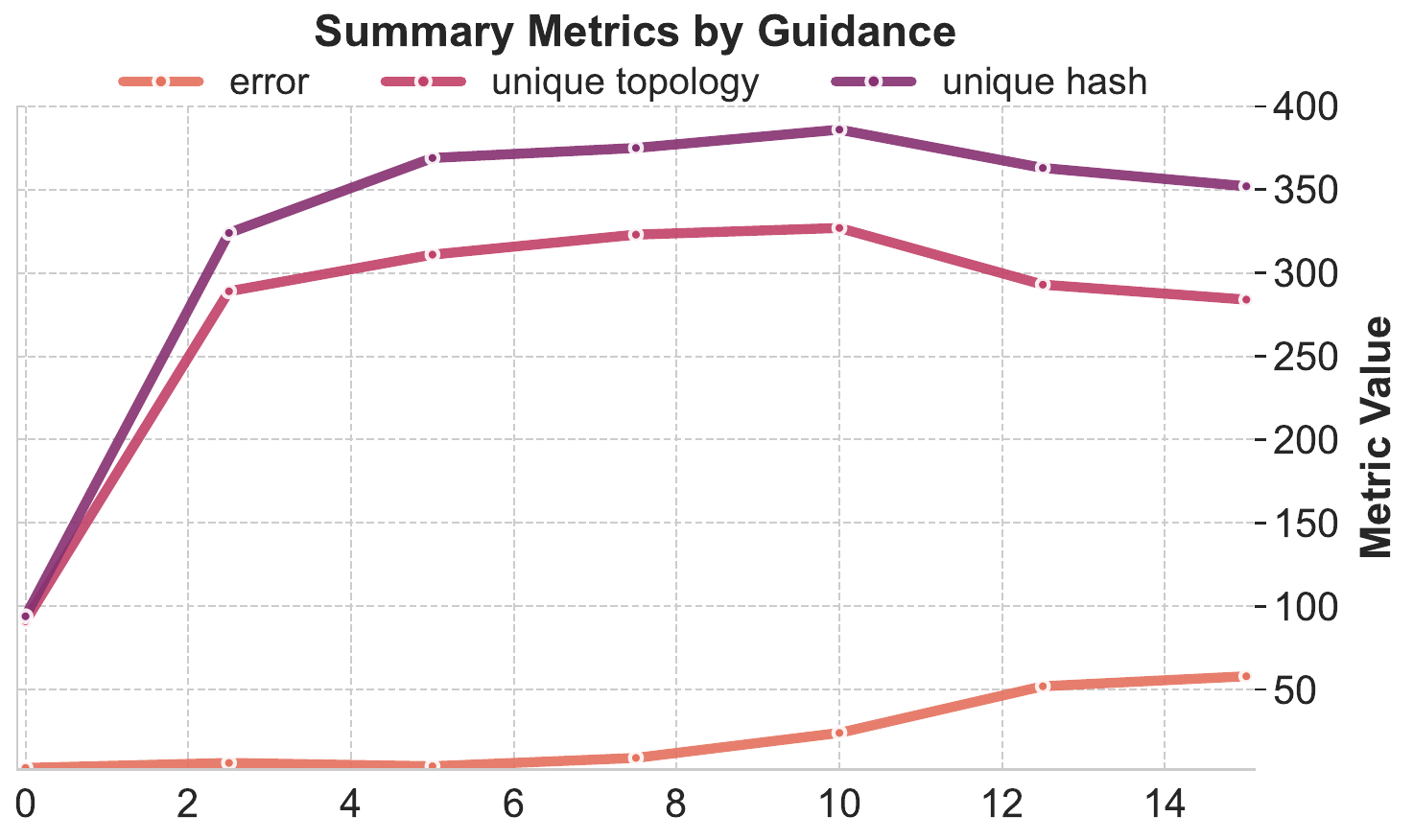}
\end{minipage}
\begin{minipage}[c]{0.49\linewidth}
\centering
\includegraphics[width=\linewidth]{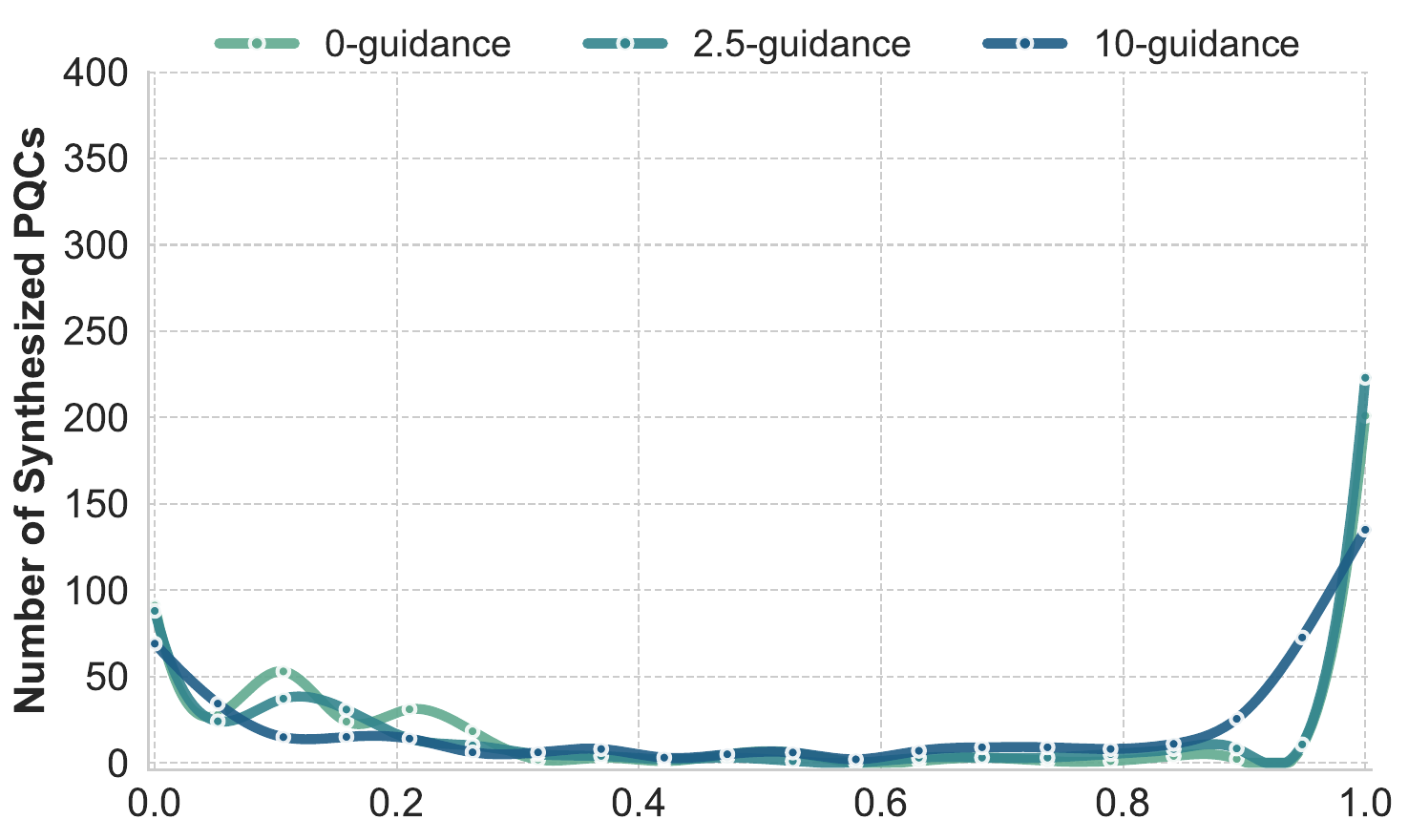}
\end{minipage}
\begin{minipage}[c]{0.49\linewidth}
\centering
\includegraphics[width=\linewidth]{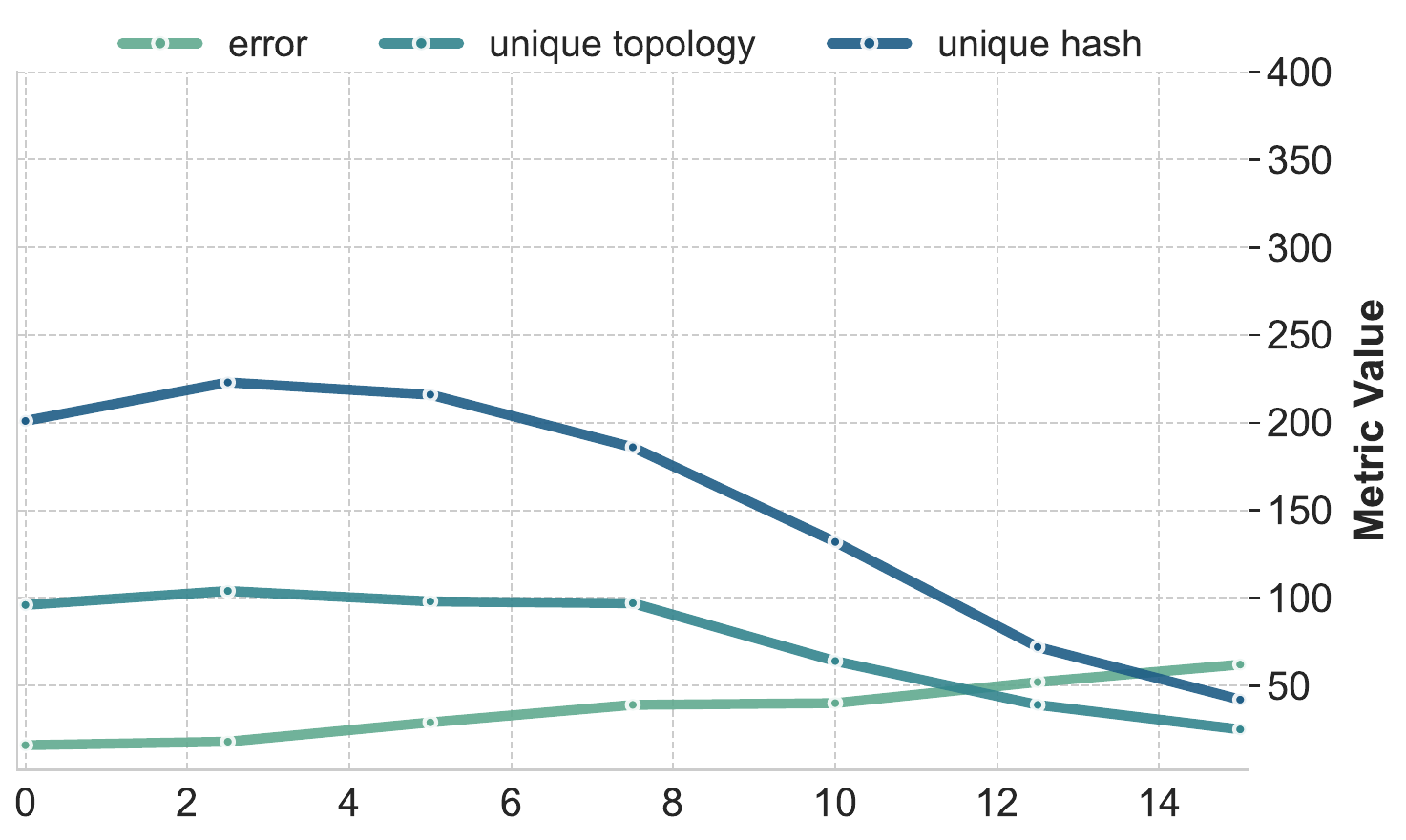}
\end{minipage}
\begin{minipage}[c]{0.49\linewidth}
\centering
\includegraphics[width=\linewidth]{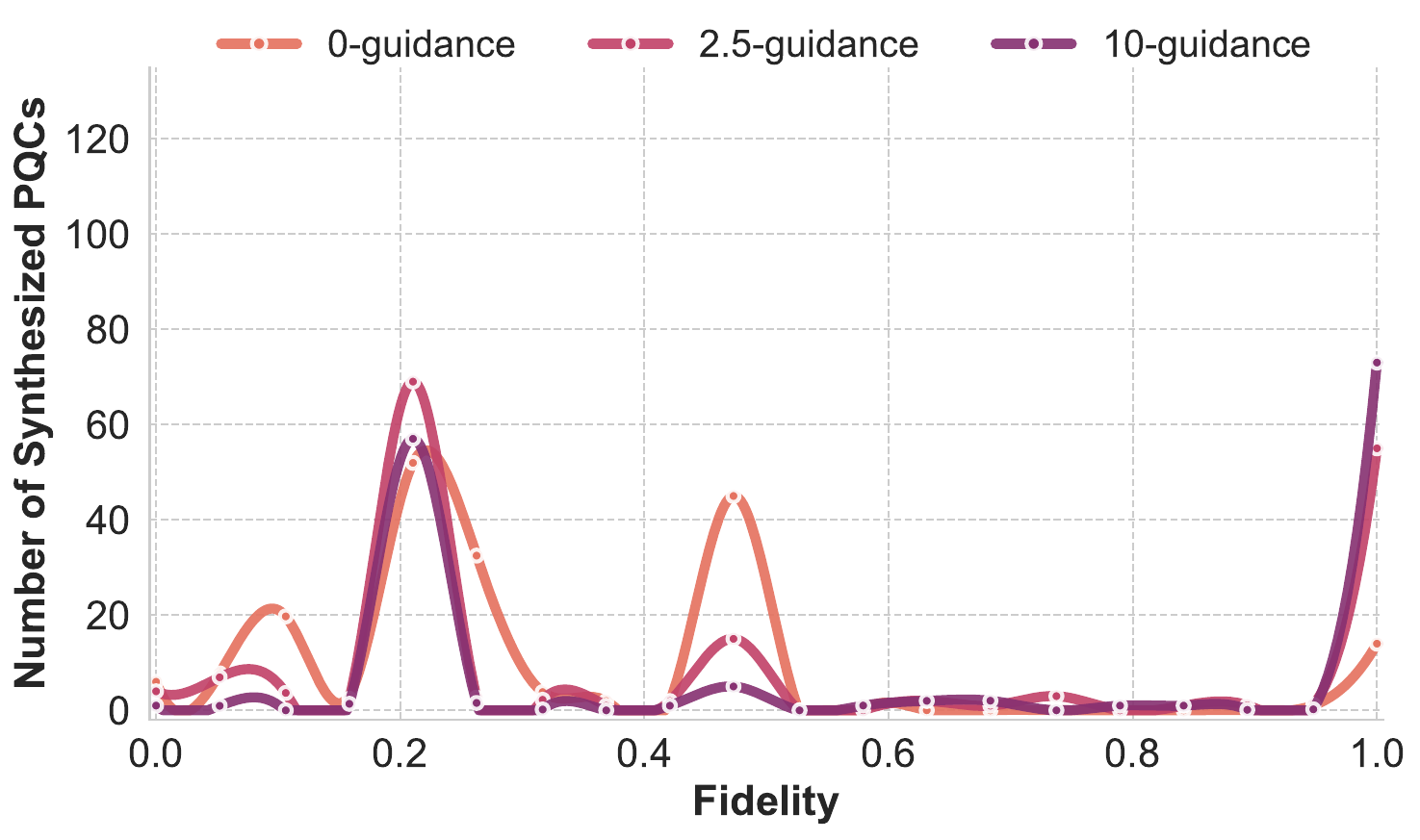}
    \end{minipage}
    \begin{minipage}[c]{0.49\linewidth}
        \centering
        \includegraphics[width=\linewidth]{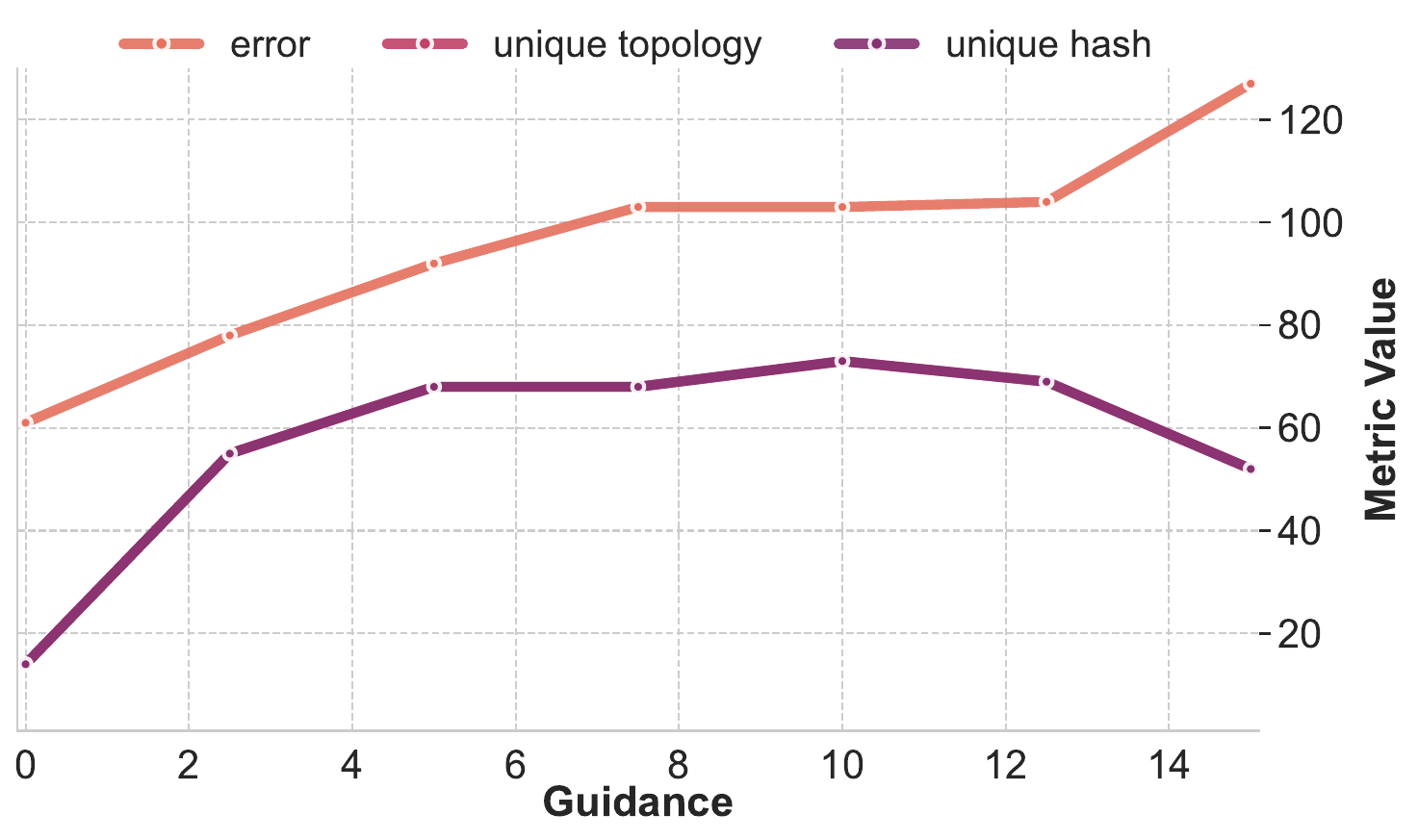}
    \end{minipage}
\caption{
\textbf{Comparison of fidelity distributions and evaluation metrics for two different gate sets: $\textbf{gs1}$ (top) and $\textbf{gs2}$ (middle) and generalization ability of our DM on 4 qubits for $\textbf{gs1}$ (bottom).}
    Shown are the fidelity scores of synthesized PQCs under varying guidance scales (left) and their respective evaluation results, e.g.\ error counts (invalid circuits) and uniqueness (right).
    }
    \label{fig:distr_eval}
\end{figure}


In this section, the results of applying our diffusion model to the GHZ state-preparation task with two distinct gate sets, $gs1$ and $gs2$, are presented. For each setting, 500 samples were synthesized to ensure statistically meaningful conclusions. As shown in Fig.~\ref{fig:distr_eval}, several key metrics are analyzed: the fidelity distributions (left), as well as evaluation metrics including error counts, structural uniqueness, and parameter uniqueness (unique hash) (right). Here, error counts denote the number of outputs from the diffusion model that could not be decoded into valid quantum circuits. Structural uniqueness quantifies the number of structurally distinct circuit architectures generated, while parameter uniqueness measures the diversity of continuous gate parameters among circuits with identical architecture, capturing variations that yield distinct quantum operations. Uniqueness metrics were evaluated only for circuits that reached the target performance. 

\begin{figure}[!b]
    \centering
    \begin{minipage}{0.49\linewidth}
        \centering
        \includegraphics[width=\linewidth]{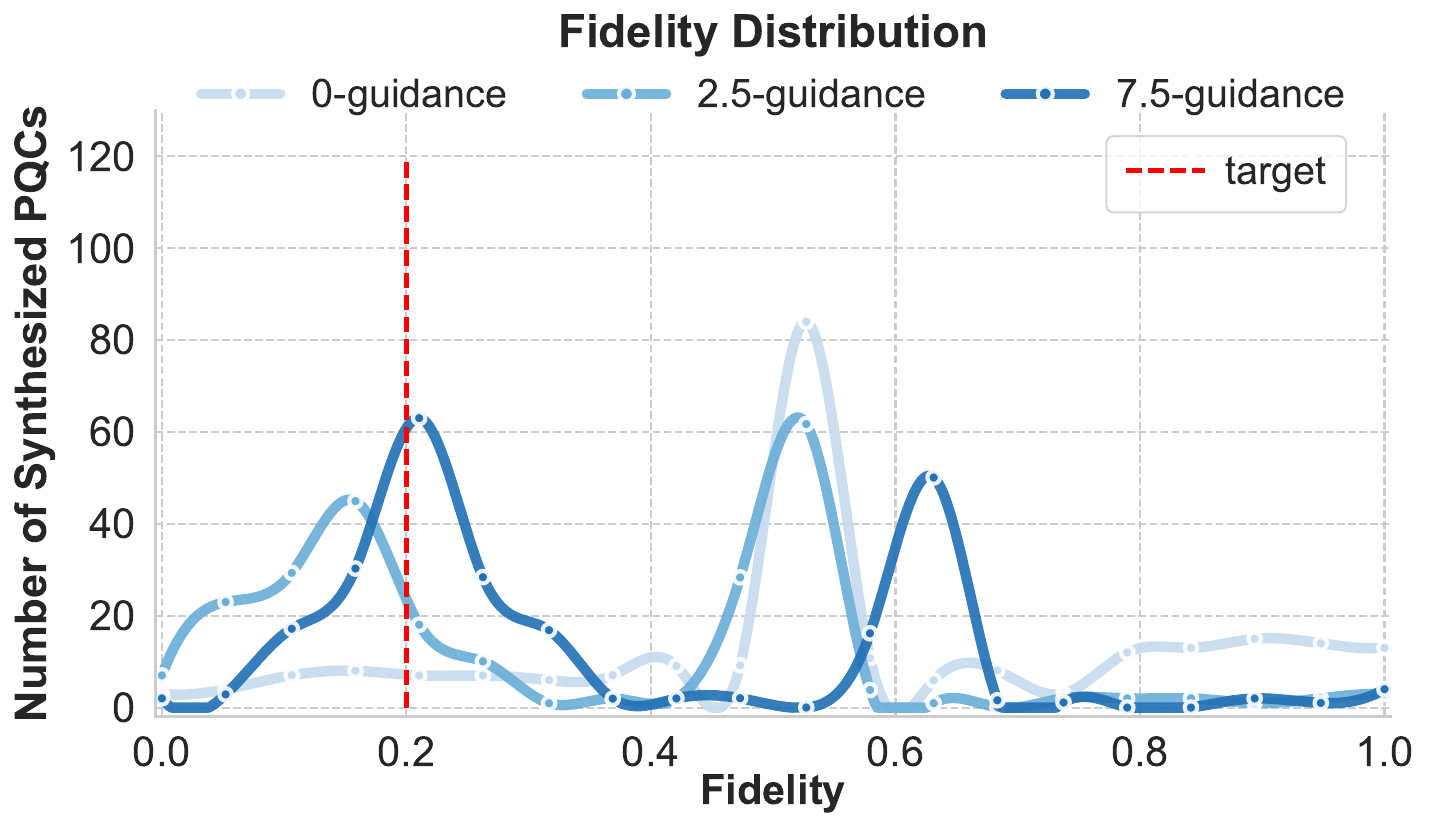}
    \end{minipage}
    \hfill
    \begin{minipage}{0.49\linewidth}
        \centering
        \includegraphics[width=\linewidth]{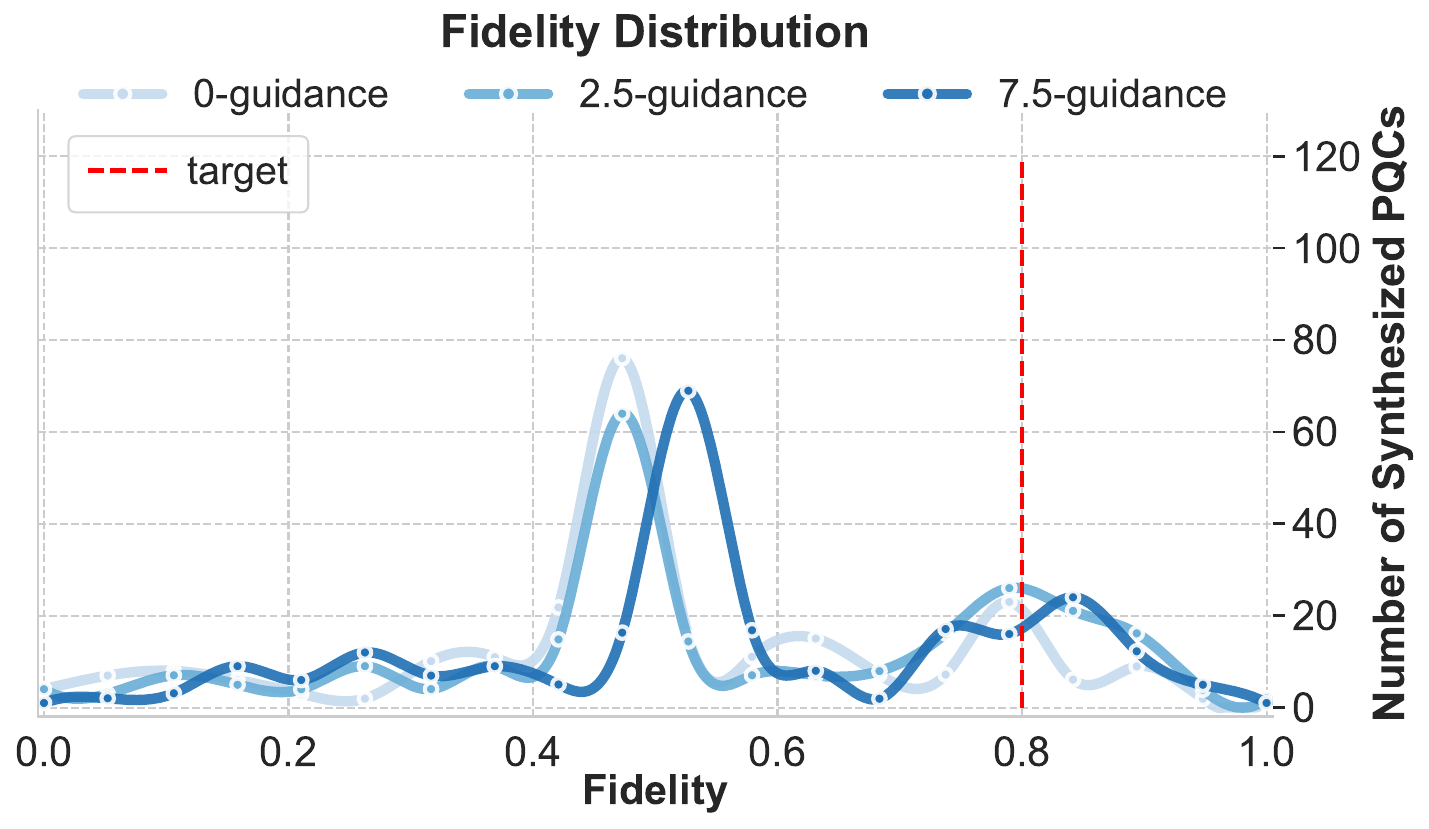}
    \end{minipage}
    \caption{\textbf{Output distribution to match different accuracy targets demonstrating the model's adaptability and controlled circuit generation}: synthesized PQCs with target accuracy of $0.2$~(left) and synthesized PQCs with target $0.8$~(right).}
    \label{fig:ml_task}
\end{figure}

In the following, the critical impact of~\gls{cfg} on circuit synthesis is discussed, demonstrating how adjusting guidance parameters modulates the generation of \gls{pqcs} tailored to the desired task.
For gate set~$gs1$~(Fig.~\ref{fig:distr_eval},~top), and with a guidance scale of $0$, the diffusion model generates only valid PQCs, each featuring a single gate per timestep and no decoding errors. The resulting fidelity distribution peaks around $0.5$. Increasing the guidance to $10$ shifts the fidelity sharply toward the target value of $1.0$, with nearly all synthesized circuits achieving perfect fidelity. Notably, even at this high fidelity, the model continues to produce a diverse array of circuit architectures; the majority of circuits exhibit structural uniqueness, and all demonstrate parameter (hash) uniqueness. This structural diversity is particularly valuable for applications requiring noise robustness or device calibration, as it enables exploration of multiple solutions. In contrast, the $gs2$ gate set (Fig.~\ref{fig:distr_eval},~middle) introduces hardware-native gates like ‘sx’ and ‘rzz’, making the synthesis task more challenging. Without CFG, the model produces circuits with fidelities spread across lower values, reflecting the increased difficulty of synthesizing with a more complex alphabet. Introducing guidance modestly increases fidelity, but the benefit plateaus quickly, reaching a maximum at a guidance scale of $2.5$. Beyond this, further increases in guidance lead to a substantial drop in both structural and parameter uniqueness, while error counts rise sharply. This highlights the added complexity and trade-offs associated with hardware-constrained synthesis tasks.

When evaluating the model's generalization ability in a zero-shot setting (Fig.~\ref{fig:distr_eval},~bottom), it is observed that a model trained exclusively on 3-qubit circuits is nevertheless able to generate valid 4-qubit circuits, despite never having encountered such configurations during training. Moreover, increasing the guidance scale enables the model to shift the fidelity distribution toward the desired target value, displaying that the DM understands how PQCs must be structured and parameterized to achieve a given task. However, this comes at the cost of a significantly increased error rate, suggesting that finetuning the model on multi-qubit data could further enhance performance. For five-qubit circuits, the diffusion model is able to produce a small number of circuits that achieve the ideal GHZ state when guidance is applied, but the error count rises substantially with increasing circuit size. These findings underscore both the scalability potential and current limitations of the approach, highlighting the value of targeted finetuning for more challenging generalization tasks.

\subsection{ML task}

Beyond GHZ-state preparation, the diffusion model shows strong performance on \gls{qml} classification tasks. To assess this, 250 circuits were generated with explicit conditioning on various accuracy targets, allowing evaluation of the model’s ability to produce PQCs that meet specific performance requirements. Although the dataset included few circuits achieving high accuracy ($\geq$ 0.7), the model reliably generated circuits clustered around the requested accuracy targets, as seen for $0.2$ (Fig.~\ref{fig:ml_task},~left) and $0.8$ (Fig.~\ref{fig:ml_task},~right), as guidance increased, while maintaining high diversity in both circuit structures and parameter values, along with a low error rate. This highlights the adaptability of the model, as it successfully transitions from generating circuits for fixed-state objectives to producing circuits that perform well on classification tasks using linearly separable data.

\subsection{Inference Efficiency and Structural Insights}

\begin{figure}[!b]
    \begin{minipage}{1\linewidth}
        \centering
        \includegraphics[width=\linewidth]{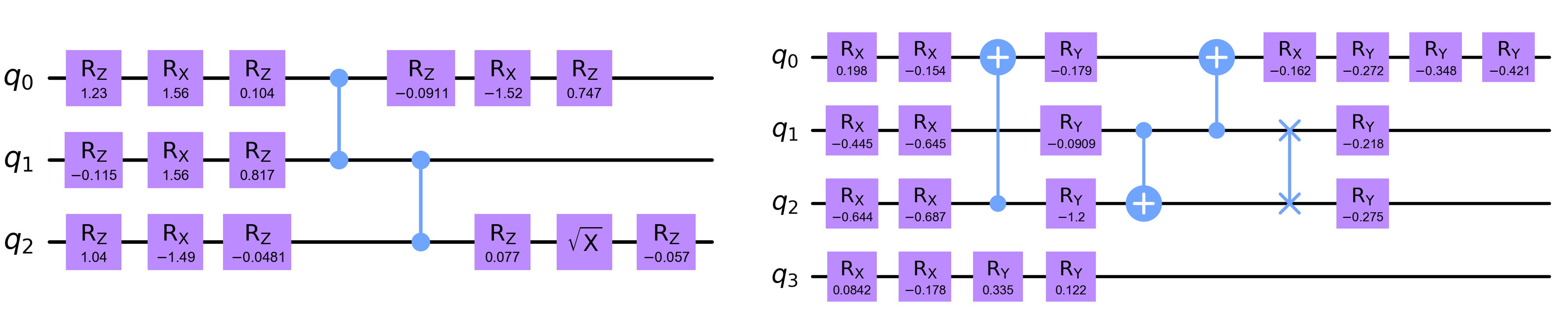}
    \end{minipage}
    \caption{\textbf{Example PQCs with layerwise gate structures generated by~DM
    :} Synthesized PQC from $gs2$ with high fidelity guidance (left) and a classification PQC with high accuracy guidance (right).}
    \label{fig:circuit}
\end{figure}

A significant discovery within this context is the autonomous emergence of structured, layer-like circuit ansätze composed of alternating single-qubit rotations and entangling gates (Fig.~\ref{fig:circuit}). The consistent appearance of these patterns suggests that the model has implicitly distilled and generalized fundamental architectural principles from the training data. Importantly, this structural regularity does not limit parameter diversity, as each layer maintains distinct continuous parameter values.
To further assess the practicality and scalability of our model, which was trained only on 3-qubit circuits with a maximum of 16 gates for the $gs1$ task, we systematically analyzed the model's inference efficiency and output quality as a function of circuit size and gate count. Tables~\ref{tab:scalability-qubit} and~\ref{tab:scalability-gates} summarize key results. Table~\ref{tab:scalability-qubit} shows the impact of increasing the number of qubits (with fixed gate count), where generation and conversion times (time needed to decode the tensor into a circuit) remain almost constant, but higher qubit counts lead to reduced numbers of synthesized high-fidelity circuits, less structural and parameter uniqueness, and a sharp rise in decoding errors. In contrast, Table~\ref{tab:scalability-gates} varies gate count (for three qubits) and reveals that the model produces the highest number of unique, target-fidelity circuits at intermediate depths (12–24 gates), while both shallow and very deep circuits show increased error rates and lower uniqueness. Together, these tables illustrate the model’s strong inference efficiency, but also highlight the practical trade-offs between scalability, diversity, and reliability in quantum circuit synthesis.

\begin{table}[htbp]
  \caption{Scalability and Inference Latency over Qubit Number for $gs1$ (Max Gates = 16, Guidance = 10)}
  \centering
  \resizebox{\columnwidth}{!}{%
    \begin{tabular}{|c|c|c|c|c|c|c|}
      \hline
      \shortstack{\textbf{Qubit}\\\textbf{Count}}
      & \shortstack{\textbf{Gen.}\\\textbf{Time (s)}}
      & \shortstack{\textbf{Conv.}\\\textbf{Time (s)}}
      & \shortstack{\textbf{High Fid.}\\\textbf{Count}}
      & \shortstack{\textbf{Uniq.}\\\textbf{Struc.}}
      & \shortstack{\textbf{Uniq.}\\\textbf{Hash}}
      & \shortstack{\textbf{Error}\\\textbf{Count}} \\
      \hline
      3  & 54.578 & 1.471 & 82 & 82 & 82 & 0  \\
      4  & 56.221 & 0.739 & 23 & 23 & 23 & 51 \\
      5  & 56.555 & 0.192 & 2  & 2  & 2  & 88 \\
      \hline
    \end{tabular}%
  }
  \label{tab:scalability-qubit}
\end{table}

\begin{table}[htbp]
  \caption{Scalability and Inference Latency over gate count for $gs1$ (Qubit Number = 3, Guidance = 10)}
  \centering
  \resizebox{\columnwidth}{!}{%
    \begin{tabular}{|c|c|c|c|c|c|c|}
      \hline
      \shortstack{\textbf{Max.}\\\textbf{Gates}}
      & \shortstack{\textbf{Gen.}\\\textbf{Time (s)}}
      & \shortstack{\textbf{Conv.}\\\textbf{Time (s)}}
      & \shortstack{\textbf{High Fid.}\\\textbf{Count}}
      & \shortstack{\textbf{Uniq.}\\\textbf{Struc.}}
      & \shortstack{\textbf{Uniq.}\\\textbf{Hash}}
      & \shortstack{\textbf{Error}\\\textbf{Count}} \\
      \hline
      8   & 41.486 & 1.733 & 60 & 60 & 60 & 4  \\
      12  & 43.195 & 1.362 & 85 & 80 & 85 & 1  \\
      16  & 54.578 & 1.471 & 82 & 82 & 82 & 0  \\
      20  & 61.176 & 1.539 & 80 & 79 & 80 & 0  \\
      24  & 61.756 & 1.419 & 79 & 79 & 79 & 1  \\
      28  & 88.816 & 1.246 & 67 & 67 & 67 & 15 \\
      \hline
    \end{tabular}%
  }
  \label{tab:scalability-gates}
\end{table}

\section{Discussion}
\label{sec:discussion}

In this work, we have demonstrated the potential of DMs as a flexible, scalable, and data-driven approach for synthesizing PQCs. By effectively addressing tasks in quantum state preparation, specifically GHZ state synthesis, as well as QML classification, our approach potentially offers significant advantages over traditional circuit synthesis methods. The experimental results confirm that diffusion-based models not only consistently synthesize valid circuits meeting targeted fidelity and accuracy (Fig.~\ref{fig:distr_eval} and Fig.~\ref{fig:ml_task}), but it also exhibits strong zero-shot generalization, as the model is able to generate PQCs that achieve the desired targets even for circuit sizes not seen during training~(Fig.~\ref{fig:distr_eval},~bottom), which indicates a genuine understanding of how circuits must be structured and parameterized to fulfill specific objectives.
Moreover, the approach offers practical control over the synthesis process through CFG, which enables users to navigate the trade-off between performance and diversity: higher guidance values promote precise, target-oriented circuit generation but may reduce diversity and error count, whereas lower guidance facilitates greater diversity in circuit generation, a valuable property in scenarios that require noise-tolerant designs or thorough calibration.
Beyond these performance aspects, our approach autonomously uncovers fundamental architectural heuristics, such as layer-like ansatz structures, consisting of single-qubit rotations followed by entangling gates (Fig.~\ref{fig:circuit}). The ability of the model to distill and reproduce these classical patterns illustrates a holistic exploration of the PQC solution space, going beyond mere interpolation between training samples.

A key practical strength of the diffusion-based approach lies in its computational efficiency during inference. The circuit generation time remains nearly constant as qubit number increases and scales slightly with gate count (Tables~\ref{tab:scalability-qubit},~\ref{tab:scalability-gates}). This strong scalability avoids the common bottleneck with search-based approaches, which typically require costly quantum simulations within optimization loops \cite{ostaszewski2021reinforcementlearningoptimizationvariational, patel2024curriculum, wang2022quantumnas, shen2023prepareansatzvqediffusion, Preskill2018QuantumCI}. However, this efficiency entails a trade-off: as system size and circuit depth grow, the quantity of high-fidelity, structurally unique circuits decline, accompanied by increased error rates. Such behavior underscores the quantum nature of the underlying problem, wherein scaling to larger circuits remains inherently challenging due to exponential growth in complexity, which can be avoided with finetuning on bigger qubit number or curriculum learning. 

While the current results are promising, several limitations persist. When generalizing to larger systems or more complex gate sets, the model encounters increased invalid outputs and decoding errors, reflecting the intrinsic complexity introduced by hardware-native constraints. The generation of circuits under restrictive hardware conditions is particularly challenging and may demand further model enhancements, broader dataset curation, or explicit integration of hardware-noise models within the training process. Moreover, assembling suitable training datasets for large-scale models, especially circuits requiring classical unitary computations, poses considerable challenges, highlighting the need for alternative conditioning methods, such as employing Hamiltonian-based representations for unitary evolutions, to improve efficiency.

\section{Conclusion and Outlook}\label{sec:conclusion}

In summary, we have introduced and evaluated a diffusion-based pipeline for generating PQCs across diverse tasks and hardware gate sets. Our empirical results underscore its capacity to deliver high-fidelity GHZ states, optimize classifiers, and flexibly integrate hardware-related design constraints. Crucially, the method requires no iterative parameter optimization or direct quantum-state simulation during training and generation, offering a scalable alternative to existing synthesis strategies.

Looking ahead, several directions remain for further development. Expanding the model to handle multi-parameter gates and leveraging starting ansätze tailored to specific quantum tasks could enhance both expressiveness and synthesis efficiency. Further research could explore larger qubit systems, integrate noise mitigation, and adapt the approach to a wider range of variational quantum algorithms. The demonstrated synergy between CFG and diffusion modeling presents opportunities for task-specific customization, including nuanced trade-offs between circuit depth, fidelity, and hardware constraints. Collectively, these advancements could further consolidate DMs as a foundation for automated quantum circuit design and accelerate progress across both academic and industrial quantum research.

\bibliographystyle{IEEEtran}
\bibliography{main} 

\end{document}

%% file: acronyms.tex
\newacronym{dms}{DMs}{diffusion models}
\newacronym{qml}{QML}{quantum machine learning}
\newacronym{nisq}{NISQ}{noisy intermediate-scale quantum}
\newacronym{ml}{ML}{machine learning}
\newacronym{pqcs}{PQCs}{parameterized quantum circuits}
\newacronym{rl}{RL}{reinforcement learning}
\newacronym{vqe}{VQE}{Variational Quantum Eigensolver}
\newacronym{qas}{QAS}{quanutm architecture search}
\newacronym{nas}{NAS}{neural architecture search}
\newacronym{gnn}{GNN}{graph neural network}
\newacronym{ghz}{GHZ}{Greenberger–Horne–Zeilinger}
\newacronym{cfg}{CFG}{classifier-free guidance}  
\newacronym{ddpm}{DDPM}{Denoising Diffusion Probabilistic Model}
\newacronym{clip}{CLIP}{Contrastive Language-Image Pre-training}